\documentclass[%
reprint,
superscriptaddress,
 amsmath,amssymb,
 aps,
 pra,
]{revtex4-2}

\usepackage{graphicx}% Include figure files
\usepackage{dcolumn}% Align table columns on decimal point
\usepackage{bm}% bold math
\usepackage{floatrow}
\usepackage{comment}
\usepackage{mathrsfs}
\usepackage{xcolor}
\usepackage{hyperref}
\usepackage{cleveref}
\crefname{section}{§}{§§}
\Crefname{section}{§}{§§}
\hypersetup{
  colorlinks   = true, %Colours links instead of ugly boxes
  urlcolor     = blue, %Colour for external hyperlinks
  linkcolor    = blue, %Colour of internal links
  citecolor   = blue %Colour of citations
}

\begin{document}

\preprint{APS/123-QED}

\title{Strong electron-phonon coupling and predicted highest known $T_{c}$ of MXenes revealed in 2H-Mo$_{2}$N under biaxial stress}

\author{Komsilp Kotmool}
\email{komsilp.ko@kmitl.ac.th}
\affiliation{College of Advanced Manufacturing Innovation, King Mongkut's Institute of Technology Ladkrabang, Bangkok 10520, Thailand}
%\affiliation{Electronic and Optoelectronic Device Research Unit, School of Science, King Mongkut's Institute of Technology Ladkrabang, Bangkok 10520, Thailand}

\author{Prutthipong Tsuppayakorn-aek}
\affiliation{Extreme Conditions Physics Research Laboratory and Center of Excellence in Physics of Energy Materials (CE:PEM), Department of Physics, Faculty of Science, Chulalongkorn University, Bangkok 10330, Thailand}

\affiliation{Thailand Center of Excellence in Physics, Ministry of Higher Education, Science, Research and Innovation, 328 Si Ayutthaya Road, Bangkok, 10400, Thailand}

\author{Thiti Bovornratanaraks}

\affiliation{Extreme Conditions Physics Research Laboratory and Center of Excellence in Physics of Energy Materials (CE:PEM), Department of Physics, Faculty of Science, Chulalongkorn University, Bangkok 10330, Thailand}

\affiliation{Thailand Center of Excellence in Physics, Ministry of Higher Education, Science, Research and Innovation, 328 Si Ayutthaya Road, Bangkok, 10400, Thailand}

\author{Thanayut Kaewmaraya}
\affiliation{Integrated Nanotechnology Research Center, Department of Physics, Khon Kaen University, Khon Kaen, Thailand}
\affiliation{Institute of Nanomaterials Research and Innovation for Energy (IN-RIE), NANOTEC-KKU RNN on Nanomaterials Research and Innovation for Energy, Khon Kaen University, Khon Kaen, 40002, Thailand}

\author{Rachsak Sakdanuphab}
\affiliation{College of Advanced Manufacturing Innovation, King Mongkut's Institute of Technology Ladkrabang, Bangkok 10520, Thailand}
\affiliation{Research Center for Quantum Technology, Faculty of Science, Chiang Mai University, Chiang Mai 50200, Thailand}

\author{Aparporn Sakulkalavek}
\affiliation{Department of Physics, School of Science, King Mongkut’s Institute of Technology Ladkrabang, Bangkok 10520, Thailand}
\affiliation{Research Center for Quantum Technology, Faculty of Science, Chiang Mai University, Chiang Mai 50200, Thailand}

\author{Rajeev Ahuja}
\affiliation{Condensed Matter Theory Group, Materials Theory Division, Department of Physics and Astronomy, Uppsala University, Box 516, SE-751 20, Uppsala, Sweden}

\affiliation{Department of Physics, Indian Institute of Technology (IIT), Ropar, Rupnagar 140001, Punjab, India}

\author{Wei Luo}
\affiliation{Condensed Matter Theory Group, Materials Theory Division, Department of Physics and Astronomy, Uppsala University, Box 516, SE-751 20, Uppsala, Sweden}

\date{\today}% It is always \today, today,
\begin{abstract}
This letter reports the unexpectedly strong electron-phonon coupling (EPC) and the highest $T_c$ record ($\approx$ 38 K) among the MXenes revealed in the 2H-Mo$_2$N under biaxial stress. At first, its excellent mechanical properties are demonstrated with ideal strength of 37 GPa and elastic modulus of 438 GPa. Subsequently, EPC and corresponding $T_c$ are elucidated upon the dynamically stable range of strain. For strain-free 2H-Mo$_2$N, the EPC constant ($\lambda$) and $T_c$ are 1.3 and 22.7 K, respectively. This $T_c$ is higher than those of 2H-Mo$_2$C (4.3 K), 1T-Mo$_2$N (16.8 K), and other pristine MXenes. The material exhibits remarkable enhancement in $\lambda$ and $T_c$ when subject to compressive and tensile stresses. The obvious strong EPC with $\lambda$ over 2.0 occurs at strains of -4{\%}, -2.5{\%}, and 5{\%}, yielding $T_c$’s of 37.8, 35.4, and 28.9 K, respectively. Our findings suggest that the strain-dependent feature and energy levels of electronic bands play an essential role in enhancing EPC. Moreover, the stronger EPC in Mo$_2$N compared with Mo$_2$C is clarified based on lattice vibrations. Therefore, this work paves a practical way for designing 2D superconducting materials using tuning atomic recipes and strain-dependent engineering.
\end{abstract}

\maketitle 

%\section{Introduction}
Superconductivity is a physical phenomenon in which the electric current can flow through a superconducting material with zero resistance below a specific superconducting critical temperature ($T_c$). The Bardeen-Cooper-Schrieffer (BCS)-type superconductors have been successfully described by the electron-phonon coupling (EPC), which creates the Cooper pairs. After MgB$_2$ was discovered by processing the  $T_c$ of 39 K\cite{R01,R02}, one of the feasible routes of hunting the high-$T_c$ conventional superconductors is to consider the two-dimensional (2D) superconducting materials. Significant discoveries such as NbSe$_2$ few layers (3 - 7 K)\cite{R03,R04}, MgB$_2$ monolayer (20 K)\cite{R05}, Mg$_2$B$_4$C$_2$ monolayer (47 K)\cite{R06}, and hydrogenated MgB$_2$ monolayer (67 K)\cite{R07} have been reported. The advantage of this path is more practical than that of the rare-earth polyhydrides, which must be operated at extremely high pressure to attain the high-$T_c$\cite{R08,R09,R10,R11,Firt}.

Mo$_2$X-based MXenes (X = C and N) have been attractive because they process many promising properties, such as visible-light photocatalysis for H$_2$ production\cite{R11-1}, thermoelectricity\cite{R12,R13}, electrochemical catalysis\cite{R14,R15,R16,R17}, and energy storage\cite{R18,R19}. Moreover, Mo$_2$C and Mo$_2$N exhibit superconducting materials with significantly high $T_c$ among the MXene members. Experimentally, 2D ultrathin $\alpha$-Mo$_2$C was successfully synthesized with $T_c$ around 3-7 K\cite{R20}. Its $T_c$ proportionally depends on sample thickness. As suggested by \textit{ab initio} calculations, two monolayers are possible, namely tetragonal (1T) and hexagonal (2H) structures. The 2H has been proposed to be more energetically favorable in both Mo$_2$C and Mo$_2$N\cite{R21,R22,R23}. These two phases differ by the position of a Mo atom as described in the Fig. 1S of Supplementary Information (SI). The  $T_c$ values of 1T and 2H of Mo$_2$C were calculated to be 7.1 K\cite{R24} and 3.2 K\cite{R25}, respectively. For instance, the $T_c$ of Mo$_2$N was theoretically reported as 16 K\cite{R24}. Interestingly, replacing C with N can enhance the $T_c$ of the 1T-Mo$_2$X more than two times. Moreover, other forms of metal nitride; for example, W$_2$N$_3$ also had been reported to be strong EPC with the $T_c$ about 22-38 K\cite{W2N3-01,W2N3-02}. These findings induce more investigations of metal nitrides, particularly the 2H-Mo$_2$N, which is more stable than another and has not yet been investigated in superconductivity.

Incorporating strain engineering into MXenes enables tuning the electronic band structures, which intrinsically changes physical properties. For example, the spin-gapless semiconductor of Ti$_2$C could be induced by biaxial strain ($\varepsilon$) over 2{\%}. Compressive strain causes the semiconductor-to-metal transitions of Hf$_2$CO$_2$ and Zr$_2$CO$_2$. The transitions of the indirect-to-direct bandgap of M$_2$CO$_2$ (M=Sc, Ti, and Cr) can be attained by applying biaxial stress. Moreover, the $T_c$ of Nb$_2$CO$_2$ is increased by 4 K under applying a biaxial tensile strain of 4{\%}. Hence, applying strain results in profound consequences on the superconductivity of MXenes. This work systematically investigates the strain-dependent $T_c$ of 2H-Mo$_2$N using density functional theory. The effect of N on EPC compared with C in Mo$_2$X is analyzed through the vibrational modes of X atoms. The structural, mechanical, and electronic properties are also calculated. This theoretical study raises the feasible route of developing MXene-based superconductors.

%\section{Computational details}
The vacuum-slab model, stress-strain response, and mechanical properties were performed by details following SI. We used the CASTEP code\cite{CASTEP} with ultrasoft pseudopotential\cite{ultrasoft} as the electronic configurations of Mo: 4s${^2}$ 4p${^6}$ 4d${^5}$ 5s${^1}$ and N: 2s${^2}$ 2p${^3}$ under GGA-PBE scheme\cite{GGA}. A dense Brillouin-zone (BZ) sampling grid of spacing 0.04$\times$2$\pi$ \AA$^{-1}$ and an E$_{cut}$ = 500 eV were verified for the sufficiently accurate geometry optimizations. The van der Waals (vdW) dispersion within Grimm’s scheme\cite{Grimme} was included in the calculations. Phonon dispersion was performed based on supercell, and finite displacement approaches with more accurate calculation (5$\times$10$^{-8}$ eV/atom and 1 meV/\AA\ for energetic and force convergences) to confirm the dynamic stability. The phonon-mediated superconductivity was investigated using the isotropic Eliashberg theory, as implemented in the QE code\cite{QE}. The E$_{cut}$ for the plane-wave basis set was 80 Ry. The EPC matrix elements were computed in the first Brillouin zone (BZ) with 8 $\times$ 8 $\times$ 1 q-meshes, while the individual EPC matrices were sampled with 64 $\times$ 64 $\times$4 k-points mesh. A Gaussian broadening parameter of 0.02 Ry was used for the integration over the Fermi surface (FS) to evaluate the electron-phonon interaction. The FS was visualized using the FermiSurfer code\cite{FS}. $T_c$ value was calculated by solving the Allen–Dynes equation\cite{Allen1972} for $\lambda$$<$1.5, and the $T_{cs}$ of strong EPC ($\lambda$$>$1.5) was obtained by a modified theory ($T_{cs}$=$f_1f_2$.$T_c$)\cite{Allen}, as also seen full detail in SI. The coulomb pseudopotential parameter ($\mu^{*}$) was set to be 0.10.

%\section{Results and discussion}
\begin{figure} 
	\centering
	\includegraphics[width=7.5 cm,angle=0]{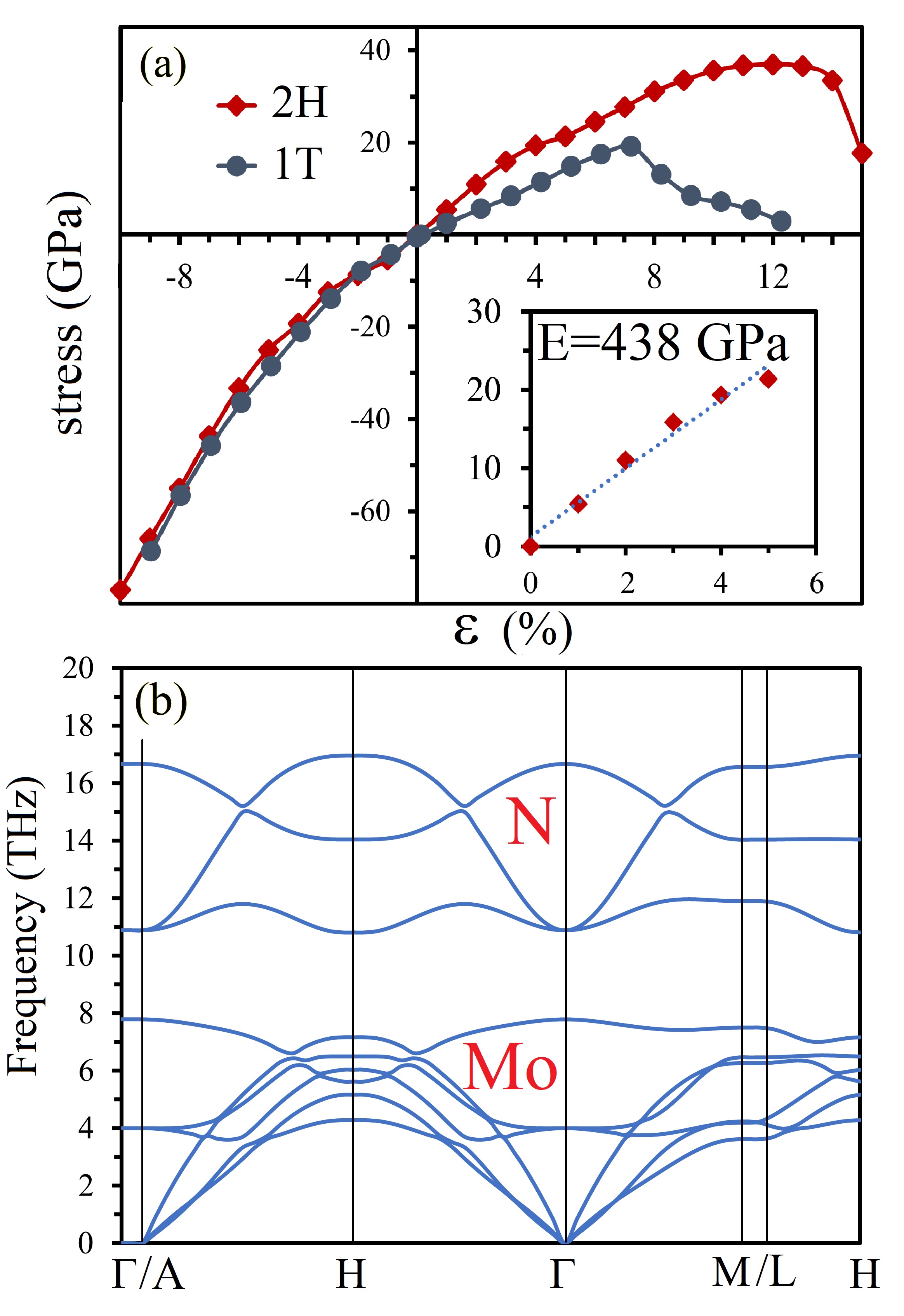}
	\caption{Stress-strain responses of the 2H-Mo$_2$N and 1T-Mo$_2$N (a), and the inset shows the linear fitting for evaluating elastic modulus. Phonon dispersion at unstressed condition (b), at compressive strains of -5{\%} and -7{\%} (c), and at tensile strains of 5{\%} and 6{\%} (d).}
	\label{Fig_01}
\end{figure}

Initially, the energetic stability of 1T- and 2H-Mo$_2$N is investigated. The 2H phase is more stable than the 1T counterpart because of having lower energy by 90 meV/atom, agreeing with the reported studies\cite{M2N}. This indicates that the 2H-Mo$_2$N is more practical than another. The 2H-Mo$_2$N has a lattice constant ($a$) of 2.818 \AA\ and a thickness ($d$) of 5.442 \AA\ at 0{\%} strain. Subsequently, the stress-strain response shown in Fig. \ref{Fig_01}a provides the ideal stress, i.e., $\sigma{_{max}}$ of 37 GPa at 12{\%} strain and elastic modulus (E) of 438 GPa, which is significantly high regarding those of the 1T-Mo$_2$N ($\sigma{_{max}}$=19 GPa at 7{\%} and E = 292 GPa). These comparative enhancements of mechanical properties of 2H-Mo$_2$N are attributed by the centrosymmetric configuration of 2H atomic layers, similar to the cases of Mo$_2$C\cite{R22} and W$_2$C\cite{R22-1}. The dynamical stability of the 2H phase is insisted in the unstressed structure as displayed in Fig. \ref{Fig_01}b. The vibrations of Mo atoms dominate at low frequencies up to 8 THz, whereas those of N atoms contribute at a higher frequency range of 11-17 THz. Stable regimes, both compressive and tensile strains, are evaluated, as it is stable in the strain range of -5{\%} to 5{\%}  (Fig. 2S). 

\begin{figure} 
	\centering
	\includegraphics[width=8.5 cm,angle=0]{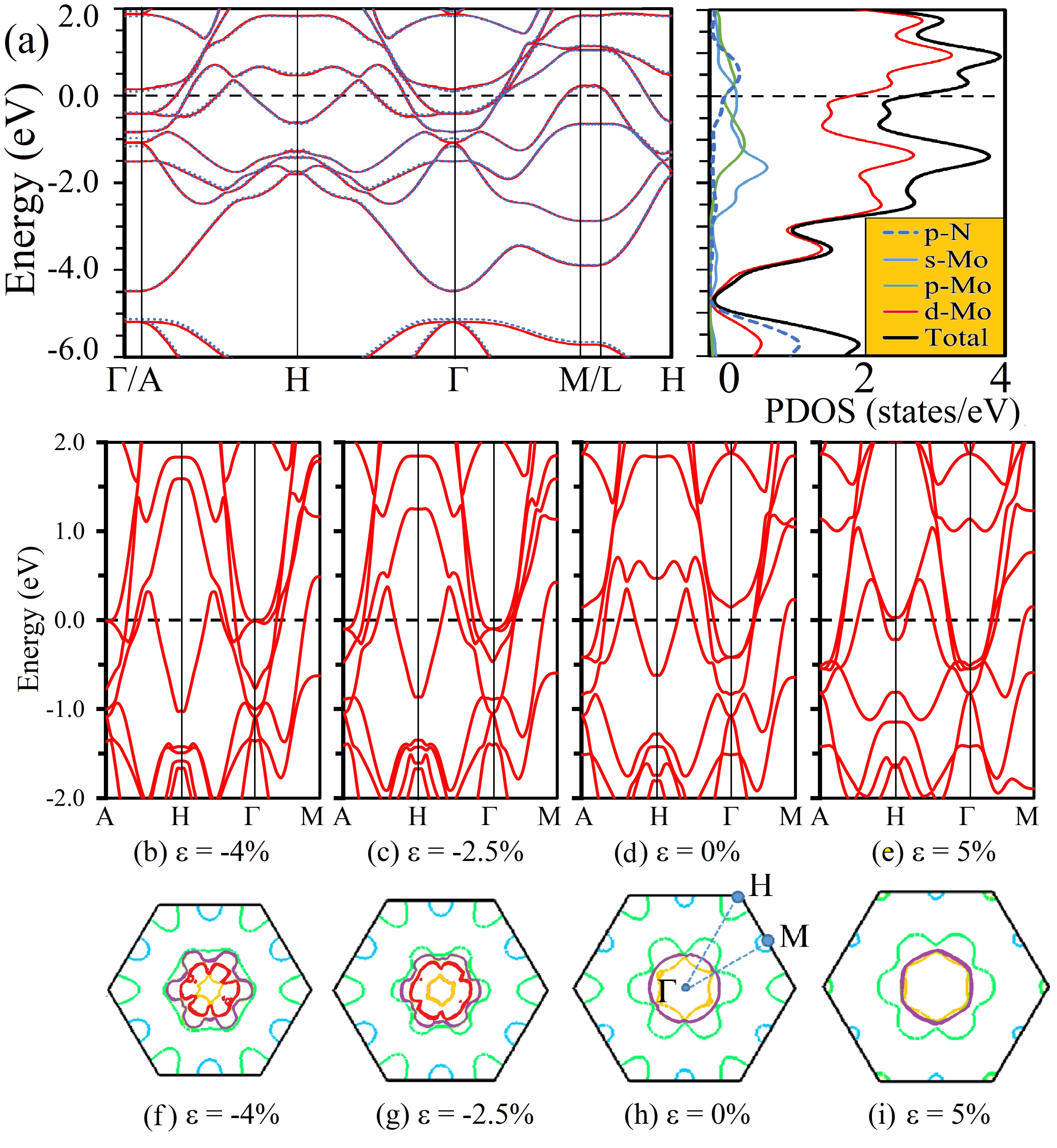}
	\caption{EBS and PDOS at (a) unstressed condition with SOC (solid-red line) and without SOC (dot-blue line), and strain-dependent EBS and FS at -4{\%} (b, f), -2.5{\%} (c, g), 0{\%} (d, h), and 5{\%} (e, i).}
	\label{Fig_02}
\end{figure}

The electronic band structure (EBS), as shown in Fig. \ref{Fig_02}a verifies the insignificant effect of spin-orbit coupling (SOC), corresponding with the previous studies in 1T-Mo$_2$N\cite{R24}. Therefore, the SOC will not be included in further calculations. The partial density of states (PDOS) reveals that around the Fermi level (FL) is mainly dominated by the 4$d$-Mo states. Moreover, minor distributions by 5$s$-Mo and 4$p$-Mo states are below the FL, and 2$p$-N states are above the FL. One can see the FL of the unstressed 2H-Mo$_2$N is located nearby the local minimum of the total DOS valley. Consequently, the FL is expected to be easily tuned to higher electronic densities. Applying stain strongly affects band structures respecting the referent bands at $\varepsilon$ = 0{\%}, as seen in Fig. \ref{Fig_02}b-e. At the tensile strain of 5{\%}, the high-energetic valence band is enhanced to higher energies at the H point, and the low-energetic conduction bands are stretched out with having the lowest touching point with the FL at the H point. Meanwhile, the valence bands at the H point decrease to lower energy at compressive strains, but those at the $\Gamma$ point are pushed up to the FL.  

To gain insights into the electronic distribution around the FL, the FS of 2H-Mo$_2$N are visualized at varying strains, as shown in Fig. \ref{Fig_02}f-i. At the reference of $\varepsilon$ = 0{\%}, FS consists of five distinctly different pockets: (i) a pancake-type electron pocket, (ii) a larger circle electron pocket, and (iii) a nest electron pocket around $\Gamma$ point, (iv) six-hole type pockets around K point, and (v) six-hole type pockets around M point, as seen in Fig. \ref{Fig_02}h. At a tensile strain of 5{\%} (Fig. \ref{Fig_02}i), pocket (i) significantly extends and nearly touches pocket (ii), and pocket (iii) climbs moderately. Moreover, pockets (iv) and (v) dramatically decline, respecting $\varepsilon$ = 0{\%}. At the compressive strains of -2.5{\%} (Fig. \ref{Fig_02}g) and -4{\%} (Fig. \ref{Fig_02}f), pocket (i) is gradually decreased, and pocket (ii) is transformed to be a nest. Interestingly, an additional nest electron pocket emerges around $\Gamma$ point, corresponding with a sinking band at this zone.  Furthermore, pocket (iii) is moderately reshaped to be a like-hexagon shape at -4{\%}. It is worth noting that the FS has one more electron pocket around $\Gamma$ at compressive strains. The magnitude of both the electron and hole pockets can be attributed to the enhancement of $T_c$ through a strain-induced electronic topological transition. Also, the band features and FS at calculated strains reflect the numbers of electron density at the FL (N$_{Ef}$) depicted in Fig. \ref{Fig_04}a. Interestingly, the N$_{Ef}$ can be rapidly enhanced by compressive strain and  gradually increased by tensile strain as a sequence of the possibility of increasing electron-phonon coupling (EPC) of this material at certain biaxial strains.

\begin{figure} 
	\centering
	\includegraphics[width=8.5 cm,angle=0]{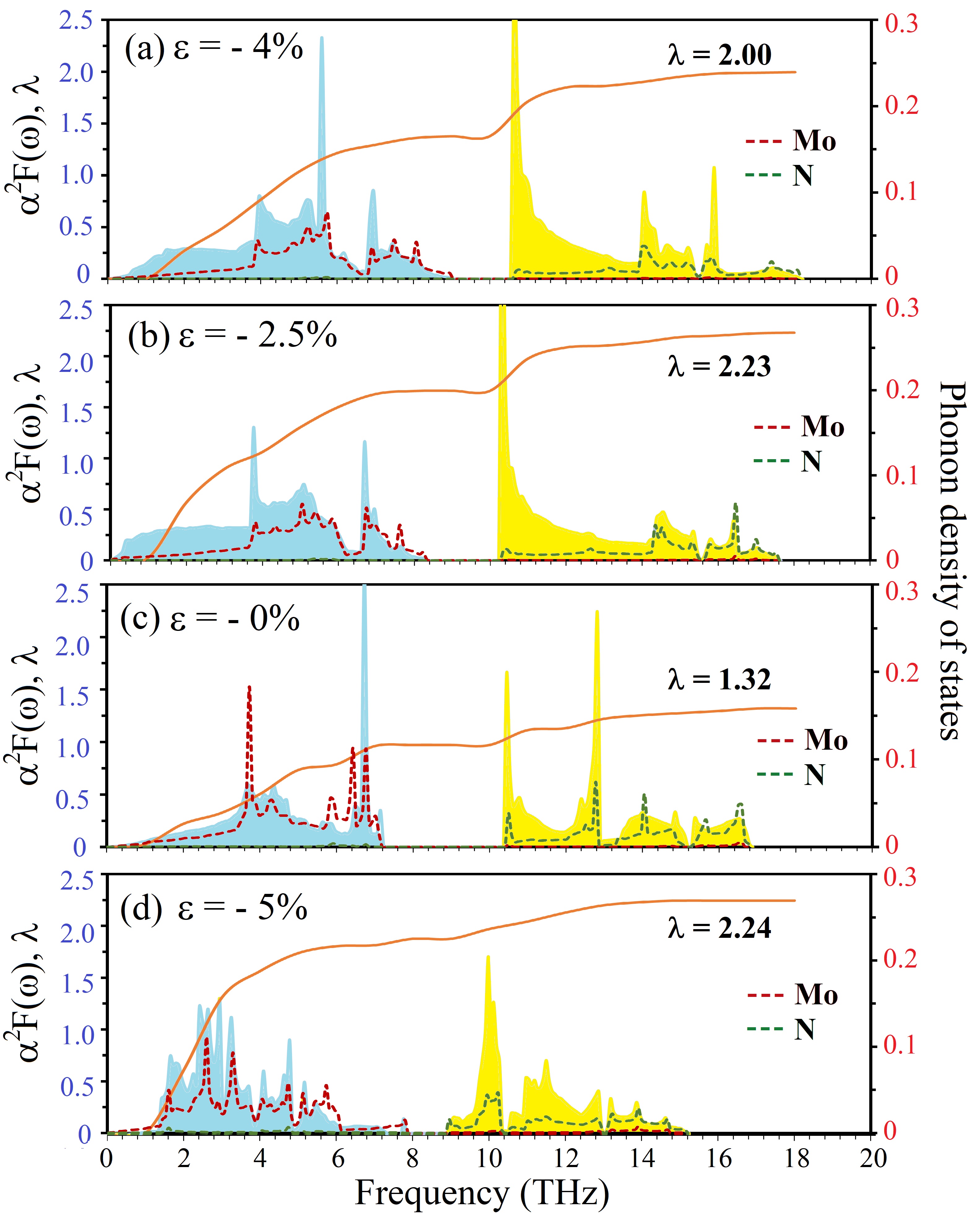}
	\caption{The $\alpha^{2}F(\omega)$(blue and yellow shades), $\lambda$ (solid orange line), and PhDOS (red and green dash lines) at -4{\%} (a), -2.5{\%} (b), 0{\%} (c), and 5{\%} (d) of 2H-Mo$_2$N. The blue and yellow shades represent the coupling modes with vibrations of Mo and N atoms, respectively. }
	\label{Fig_03}
\end{figure}

The $T_c$ values of 2H phase versus biaxial strains are calculated using the Allen-Dynes formula. This formula needs the input parameters, including the integration of the lambda ($\lambda$) and the logarithmic average of the spectral function ($\omega_{log}$), which are solved by the isotropic Eliashberg theory, as obeyed in SI.
%by equations \ref{eq:Allen-Dynes} to \ref{eq:omega}.

%\begin{equation}\label{eq:Allen-Dynes}
%T_{c} = \frac{\omega_{log}}{1.2} \exp \Big[ -%\frac{1.04(1+\lambda)}{\lambda-\mu^*(1+0.62\lambda)} \Big],
%\end{equation}

%\begin{equation}\label{eq:lambda}
%\lambda = 2 \int_{0}^{\infty}\frac{d\omega}{\omega}\alpha^{2}F(\omega),
%\end{equation}

%\begin{equation}\label{eq:omega}
%\omega_{log} = \exp \Big [\frac{2}{\lambda}\int_{0}^{\infty}\frac{d\omega}{\omega}{\alpha^{2}F(\omega)}{\ln\omega} \Big],
%\end{equation}

The patterns of Eliashberg spectral function ($\alpha^{2}F(\omega)$) of 2H-Mo$_2$N at calculated strains have a gap between the spectra contributed by vibrations of Mo and N atoms. The frequency distribution of $\alpha^{2}F(\omega)$ depends on the interaction between atoms, stretching out to higher frequency from tensioned to compressed structures. At stain-free condition, it has the widest gap and has a $\lambda$ of 1.32, as displayed in Fig. \ref{Fig_03}c. We demonstrate the enhancement of EPC of 2H-Mo${_2}$N regarding 2H-Mo${_2}$C ($\lambda$=0.58) through the distribution of the phonon linewidth ($\gamma_{\textbf{q}\nu}$) on the phonon dispersion as shown in Fig. 3S. We found that higher $T_c$ of 2H-Mo${_2}$N is interplayed by its stronger $\gamma_{\textbf{q}\nu}$ at A$^\prime$$_1$ mode (symmetrical bend) of Mo atom, E$^\prime$ mode (asymmetrical stretch) of coupling of Mo and N atoms, and A$^{\prime\prime}$$_1$ mode (asymmetrical bend) of N atom. Moreover, according to equation 8S, because the lower frequency range of N respecting C atom, it results in stronger $\lambda_{\textbf{q}\nu}$ in 2H-Mo${_2}$N. Intriguingly, Fig. \ref{Fig_03}a, \ref{Fig_03}b, and \ref{Fig_03}d demonstrate that the $\lambda$ values of 2H-Mo$_2$N at -4{\%}, -2.5{\%}, and 5{\%} strains are 2.00, 2.23, and 2.24, respectively. Vibrations of Mo atoms mainly dominate the $\lambda$ by over 70{\%}. In comparison, the rest contribution belongs to the N atom with significant enhancement at E$^\prime$ mode presented at high-frequency region (yellow shade). These $\lambda$ values are obviously classified as strong EPC behavior ($\lambda$>1.5). This finding is the first theoretical study that uncovers the existence of strong EPC in the pristine MXene family. Full details of the relative parameters of EPC are listed in Table 1S.

\begin{figure} 
	\centering
	\includegraphics[width=7.5 cm,angle=0]{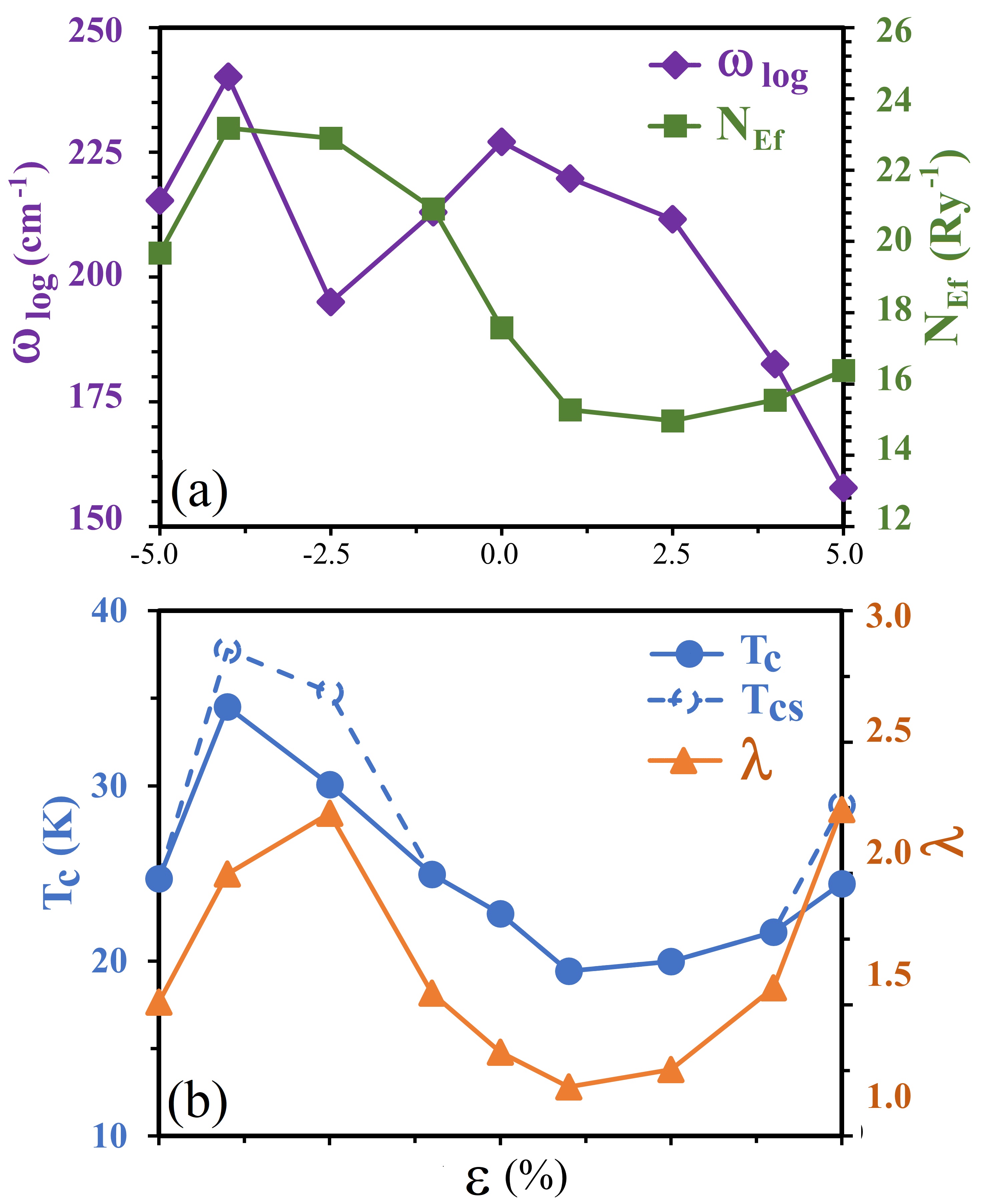}
	\caption{(a) The $\omega_{log}$ and N$_{Ef}$, and (b) the $T_c$, $T_{cs}$, and $\lambda$ of 2H-Mo$_2$N as a function of strain.}
	\label{Fig_04}
\end{figure}

Here we plot the related superconductive parameters as a function of strain in Fig. \ref{Fig_04}a-b to clarify their strain dependencies. At 0{\%} strain, the $\omega_{log}$ and $T_c$ of 2H-Mo$_2$N are 227 cm$^{-1}$ and 22.7 K, respectively. This $T_c$ value is higher than the $T_c$ of 1T-Mo$_2$N ($T_c$ = 16.8 K with $\omega_{log}$=248 cm$^{-1}$ and $\lambda$=0.98) that is calculated in this work to validate the result of previous work ($T_c$=16.0 K)\cite{R24}. Furthermore, the $T_c$ of the 2H-Mo$_2$N is extremely higher than that of 2H-Mo$_2$C ($T_c$=4.3 K), as discussed before. Therefore, it is essential to mention that the $\lambda$ and $T_c$ of bare 2H-Mo$_2$N are significantly higher than other MXenes without surface functionalization and applying external stresses. At -2.5{\%} and 5{\%} strains, the $\omega_{log}$ inversely relates with the N$_{Ef}$ which can be described by the connection of equations 5S and 7S. Moreover, it is found that the $\omega_{log}$ decreases with increasing $\lambda$, which is explicitly explained by equation 5S. The $T_c$ values of the 2H-Mo$_2$N solved by equation 3S are 34.5, 30.1, and 24.4 K for strains of -4{\%}, -2.5{\%}, and 5{\%}, respectively. According to the strong EPC at these strain values, it is important to include the correction factors ($f_1$ and $f_2$) for obtaining a more plausible superconducting temperature ($T_{cs}$=$f_1f_2$.$T_c$) as obeyed in SI. We find that the $T_{cs}$ significantly increases by 9{\%} at $\varepsilon$=-4{\%} and 18{\%} at -2.5{\%} and 5{\%} strains, as shown in Fig. \ref{Fig_04}b. The quantities of the correction and other related factors are also shown in Table 2S. Unexpectedly, the $T_{cs}$ values are 37.8, 35.4, and 28.9 K for strains of -4{\%}, -2.5{\%}, and 5{\%}, respectively.

%\section{Conclusion}
In summary, this letter reveals an unexpectedly strong EPC in 2H-Mo$_2$N under biaxial stress. The 2H-Mo$_2$N is considered because it is more energetically stable than another structure, 1T-Mo$_2$N. It processes to be a dynamically stable structure in the wide strain range of -5{\%} to 5{\%}. The application of stress plays an essential role in tuning both the feature and energy level of electronic band structure (EBS), consequently enhancing EPC. A high record $T_{cs}$ of free-strain 2H-Mo$_2$N of 22.7 K is reported. The EBS, FS, and N$_{Ef}$ are carefully elucidated to advocate a strong EPC. The corrected critical temperature for strong EPC, $T_{cs}$, is evaluated as the consequence of increasing superconducting temperatures by 9 - 18{\%}. Our finding provides a new record for the highest $T_c$ among the MXene family. This result challenges further experiments to prove it and paves the guidance for developing the MXene-based superconductor for modern material applications.

%\begin{acknowledgments}
This work is supported by the King Mongkut’s Institute of Technology Ladkrabang (KMITL) under the Fundamental Fund (grant number: KRIS/FF65/25). The computing facilities provided by the College of Advanced Manufacturing Innovation (AMI), SNIC (2021/1-42) and SNAC of Sweden are acknowledged for providing part of the computing facilities. The authors also gratefully acknowledge the Swedish Research Council (VR-2016-06014 \& VR-2020-04410) for financial support.
%\end{acknowledgments}

%\bibliographystyle{unsrt}
%\bibliography{manuscript.bib}

\end{document}